\documentclass[conference]{IEEEtran}
\IEEEoverridecommandlockouts
\usepackage{amsmath}
\usepackage{cite}
\usepackage{amssymb}
\usepackage{array}
\usepackage{amsfonts}
\usepackage{algorithm}
\usepackage{algpseudocode}
\usepackage{comment}
\usepackage[letterpaper, top=1.9cm, bottom=2.54cm, left=1.6cm, right=1.6cm]{geometry}
\usepackage{dsfont}
\usepackage{graphicx}
\usepackage{epsfig}
\usepackage{subfigure}
\usepackage{float}
\usepackage{epstopdf}
\usepackage{psfrag}
\usepackage{xcolor}
\usepackage{url}

\usepackage{mathtools}

\title{Reconfigurable Intelligent Surface-Assisted Multiuser Tracking and Signal Detection in ISAC}

\author{
\IEEEauthorblockN{Weifeng Zhu, Junyuan Gao, Shuowen Zhang, and Liang Liu}
\IEEEauthorblockA{Department of Electrical and Electronic Engineering, The Hong Kong Polytechnic University}
\IEEEauthorblockA{Email: \{eee-wf.zhu, junyuan.gao, shuowen.zhang, liang-eie.liu\}@polyu.edu.hk}
\vspace{-1cm}
\thanks{The work was supported by the Research Grants Council, Hong Kong, China, with Grant No. PolyU C5002-23Y.}
}

\begin{document}
		\maketitle 

\abovedisplayskip=1pt
\belowdisplayskip=1pt
\allowdisplaybreaks

\begin{abstract}
This paper investigates the multiuser tracking and signal detection problem in integrated sensing and communication (ISAC) systems with the assistance of reconfigurable intelligent surfaces (RISs). Due to the diverse and high user mobility, the tracking and signal detection performance can be significantly deteriorated without choreographed user state (position and velocity) updating principle. To tackle this challenge, we manage to establish a comprehensive probabilistic signal model to characterize the interdependencies among user states, transmit signals, and received signals during the tracking procedure. Based on the Bayesian problem formulation, we further propose a novel hybrid variational message passing algorithm for the online estimation of user states, which can iteratively update the posterior probabilities of user states during each tracking frame with computational efficiency. Numerical results are provided to demonstrate that the proposed algorithm can significantly improve both of the tracking and signal detection performance over the representative Bayesian estimation counterparts.
\end{abstract} 


\section{Introduction}

The advent of the sixth-generation (6G) wireless networks marks a significant revolution to encompass advanced environmental sensing with communication services under the common cellular infrastructure through integrated sensing and communication (ISAC) \cite{Liu_2022_CST,isac_survey1,isac_survey2}. As essential foundations in 6G ISAC systems, the high-precision localization and tracking capability is considered to promote various intelligent applications, including traffic monitoring, autonomous driving, and drone swarm control \cite{isac_survey1,isac_survey2,Cui_2021_NM}, which benefits from advanced multiple-input multiple-output (MIMO) and broadband signal processing techniques \cite{Stuber_2004_Proceeding}. However, line-of-sight (LOS) paths between base stations (BSs) and users are indispensable for existing localization and tracking schemes, which restricts their application for realizing seamless sensing in the practical environment with various obstacles.

To tackle the LOS blockage in ISAC systems, the reconfigurable intelligent surface (RIS) is considered as a promising solution to provide substantial sensing coverage enhancement by benefiting from its low cost and energy efficiency \cite{Chepuri_2023_SPM}. Specifically, the RIS can be deployed at a proper position to provide additional cascaded LOS channels between users and BSs, and serves as a passive anchor for localization with extra position-related measurements. Several recent works \cite{Han_2022_JSTSP,Rahal_2024_JSTSP,Zhu_2025_arxiv} have explored the possibility of localization with the assistance of RIS in ISAC systems and propose a series of localization schemes. In \cite{Han_2022_JSTSP,Rahal_2024_JSTSP}, the maximum likelihood estimation (MLE)-based localization schemes are proposed to directly estimate the distance and angle of arrival (AOA) towards the RIS from the received signal, which suffers from prohibitive complexity in the multiuser case. Meanwhile, the work \cite{Zhu_2025_arxiv} proposes an efficient subspace-based localization algorithm to localize mixed near-field and far-field users. Though these works provide useful insights for the RIS-assisted localization scheme design and analysis, the research on the RIS-assisted tracking problem with multiple high-mobility users is still in fancy.

In contrast to the localization problem, multiuser tracking belongs to the online estimation problem, which aims to realize consistently accurate position estimation during a long time interval. Although conventional localization schemes \cite{Han_2022_JSTSP,Rahal_2024_JSTSP,Zhu_2025_arxiv} can still be utilized in each tracking frame, they neglect the temporal correlation of user states (positions and velocities) during the tracking procedure. Under these methods, the tracking performance can suffer from significant degradation when the receive signal strength is weak. On the other hand, it is also intractable to directly apply the maximum \emph{a posterior} (MAP) estimation for the multiuser tracking problem due to the high-dimensional search complexity. Recently, the work \cite{Teng_2023_JSAC} models the user trajectory with a stationary Gaussian process and proposes an online Bayesian tracking algorithm. However, such modeling does not accommodate the non-stationary velocity component and can only apply to the low-mobility scenario, which suffers from severe tracking deviations in the harsh scenario with possibly high-mobility users. This work aims to address the challenging multiuser tracking problem with diverse and high-speed mobility in the RIS-assisted ISAC system. Under the pilot-free transmission framework, the BS performs joint estimation of user states and transmit signals from received signals, which can further enhance spectral efficiencies. Specifically, we first manage to establish a comprehensive probabilistic signal model for the considered system. Next, we propose a novel hybrid variational message passing (HVMP) algorithm to realize the joint estimation task with computational efficiency during the tracking interval. Numerical results reveal that the proposed algorithm can achieve significantly enhanced tracking and signal detection performance over existing Bayesian estimation benchmarks by fully exploiting statistical dependencies in user states.

\section{System Model and Problem Formulation}\label{sec:system_model}

\begin{figure}
  \centering
  \includegraphics[width=.48\textwidth]{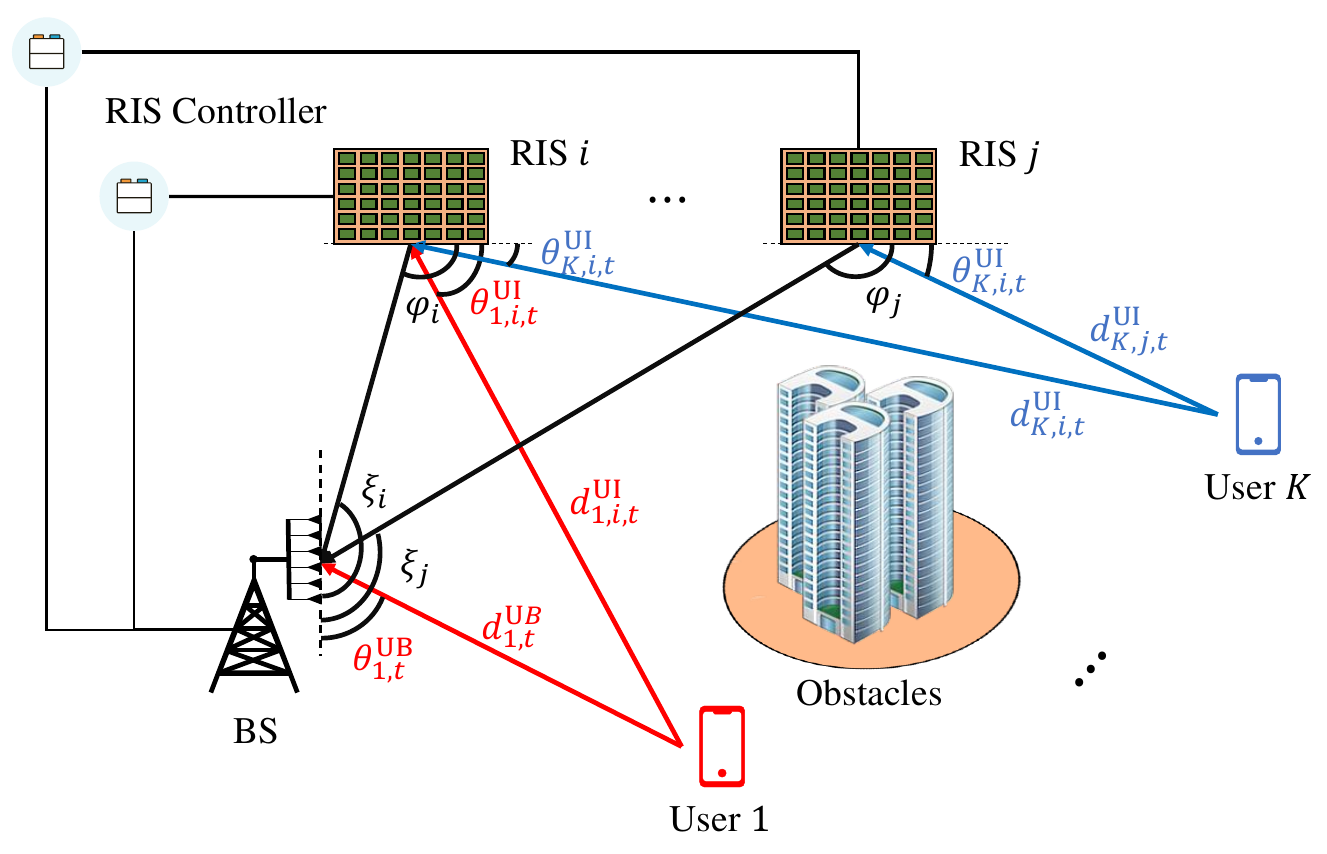}
  \vspace{-0.5cm}
  \caption{Illustration of the system model. Multiple RISs are deployed in the system to assist the sensing and communication service of users.}\label{fig:system model}
  \vspace{-0.5cm}
\end{figure}

We consider an uplink ISAC system with a multi-antenna BS and $K$ single-antenna users with mobility. Each user transmits orthogonal frequency division multiplexing (OFDM) signals to the BS in the uplink, and then the BS decodes the data and tracks the user positions based on the received signals. Due to various obstacles in the environment, we consider that LOS paths between users and BS can possibly be obstructed. To tackle this challenge, $R$ RISs equipped with $M_\text{I}$ reflecting elements are deployed at proper locations with LOS paths to the BS and users, as shown in Fig. \ref{fig:system model}.
In the two-dimensional (2D) system, the coordinate of the BS is denoted as $\mathbf{p}^{\rm B} = [p^{\text{B}}_{x},p^{\text{B}}_{y}]^T \in \mathbb{R}^{2 \times 1}$ and the coordinate of each RIS is denoted as $\mathbf{p}^{\rm I}_{r} = [p^{\text{I}}_{x,r},p^{\text{I}}_{y,r}]^T \in \mathbb{R}^{2 \times 1}$, $\forall r \in \{1,\dots,R\}$. In each frame $t$, we consider that the position of each user $k$ is denoted as $\mathbf{p}^{\rm U}_{k,t} = [p^{\text{U}}_{x,k,t},p^{\text{U}}_{y,k,t}]^T \in \mathbb{R}^{2 \times 1}$, $\forall k,t$ and the velocity is given by $\dot{\mathbf{p}}^{\rm U}_{k,t} = [\dot{p}^{\text{U}}_{x,k,t},\dot{p}^{\text{U}}_{y,k,t}]^T \in \mathbb{R}^{2 \times 1}$, $\forall k,t$. 

\subsection{Transmission Framework}

In this work, the tracking procedure contains overall $V$ consecutive frames, where $Q$ OFDM symbols in each frame are utilized for the tracking task. The tracking interval between two adjacent tracking frames is defined as $\Delta T$. Moreover, we assume that there are $N$ OFDM sub-carriers with sub-carrier spacing $\Delta f$ Hz, leading to the bandwidth being $B = N\Delta f$ Hz. 
To avoid the inter-symbol interference, a cyclic prefix (CP) comprised by $J \le L_{\rm max}$ OFDM samples is inserted at the beginning of the transmit OFDM symbol, where $L_{\rm max} \le N$ is the maximal channel tap of all users. As such, the period of each OFDM symbol is calculated by $\Delta t = \frac{N+J}{N \Delta f}$. 

\begin{figure}
  \centering
  \subfigure[Transmission frame structure for uplink users.]{
  \includegraphics[width=.45\textwidth]{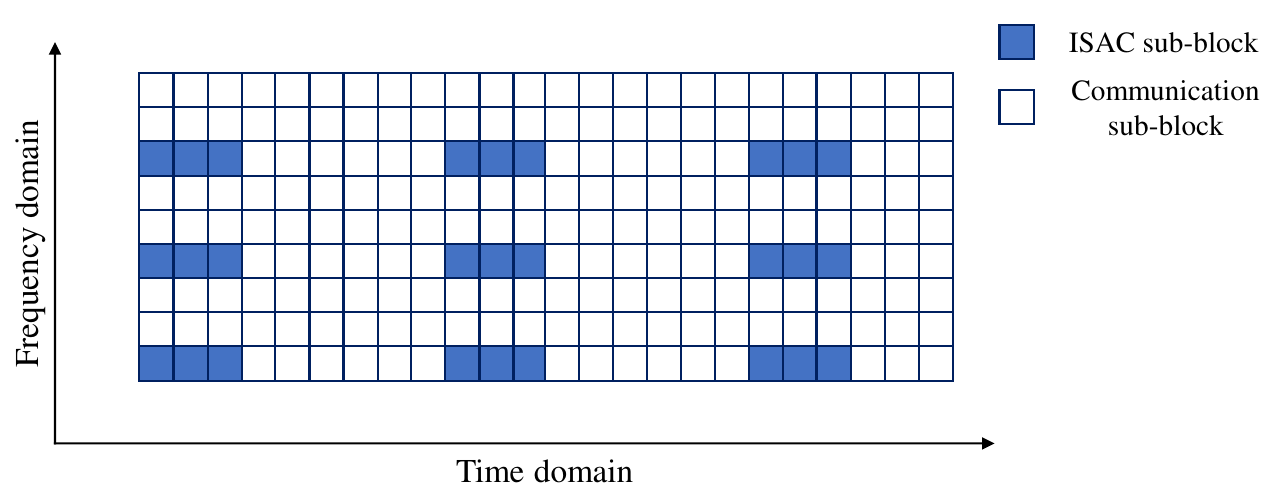}}
  \subfigure[Reflection phase profile for RIS.]{
  \includegraphics[width=.45\textwidth]{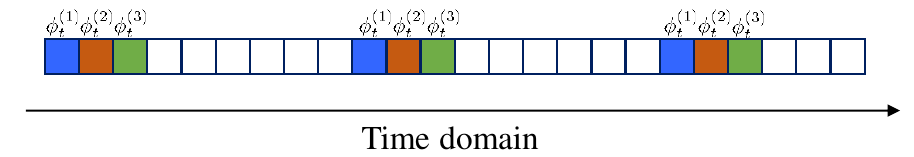}}
  \vspace{-0.3cm}
  \caption{Illustration of the transmission frame structure.}\label{fig:frame}
  \vspace{-0.7cm}
\end{figure}

We adopt a two-phase uplink transmission framework in the considered system, which consists of the ISAC sub-block and the communication sub-block. In particular, the BS accomplishes dual functionalities of sensing and communication in the ISAC sub-block. For convenience, we denote $\mathcal{Q}$ and $\mathcal{N}$ as the OFDM symbol set and subcarrier set of the ISAC sub-block. In the ISAC sub-block of each frame $t$, we adopt the repetition coding scheme with
\begin{align}
    s_{k,n,t}^{(q)} &= \tilde{s}_{k,t}, ~\forall k,t, q \in \mathcal{Q}, n \in \mathcal{N},
\end{align}
to boost the tracking accuracy.
As shown in Fig. \ref{fig:frame}, we divide the OFDM symbols in the ISAC sub-blocks into $Q_2$ groups to facilitate the AOA and Doppler frequency estimation, which contains $Q_1$ consecutive OFDM symbols. The OFDM set can be re-expressed by $\mathcal{Q} = \bigcup_{i = 1}^{Q_2} \mathcal{Q}_i$ with $\mathcal{Q}_i = \{1+(i-1)\Delta Q,\dots, Q_1+(i-1)\Delta Q\}$. Similarly, the selected $N_0$ sub-carriers in the ISAC sub-block are equally spaced with the sub-carrier set $\mathcal{N} = \{1,\dots,1+(N_0-1)\Delta N\}$. For the RIS, we follow our previous work \cite{Zhu_2025_arxiv} to design a spatiotemporal phase profile $\pmb{\Psi}_r = [\pmb{\phi}_{r,t}^{(1)},\dots,\pmb{\phi}_{r,t}^{(Q_1)}]$ which is common in each symbol group $\mathcal{Q}_i$ for each RIS $r$ to facilitate the AOA estimation for user-RIS links.
For the communication sub-block, we consider that transmit signals are independent across different sub-carriers and OFDM symbol intervals to boost the transmission rate performance, which can be successfully detected based on estimated channels in the ISAC sub-block. In this work, we concentrate on resolving the joint multiuser tracking and signal detection problem with received signals in the ISAC sub-block.


\subsection{Signal Model}

Define $\mathbf{G}_{r,n} = \beta^{\rm IB}_{r} e^{-\jmath 2 \pi \Delta f (n-1) \tau^{\rm IB}_{r}} \mathbf{a}_{\rm B}(\xi_{r}) (\mathbf{a}_{\rm I}(\varphi_{r}))^T \in\mathbb{C}^{M_{\text{B}}\times M_{\text{I}}}$ as the channel from the RIS $r$ to the BS at the $n$-th subcarrier, where $\beta^{\rm IB}_{r}$, $\tau^{\rm IB}_{r}$, $\xi_{r}$ and $\varphi_{r}$ are the complex channel gain, delay profile, AOA, and angle of departure (AOD) of the BS-RIS $r$ channel, respectively. Here, we assume that the complex channel gain depends on the distance between the BS and RIS $r$. Consider that the BS and RISs are all equipped with the uniform linear array (ULA) with $d=\frac{\lambda}{2}$ spacing, the steering vector for the BS and RISs can be expressed by
\begin{align}
    \mathbf{a}_{\rm B}(\theta) &= \big[ 1, e^{\jmath \omega_{\rm S} \cos \theta}, \dots, e^{\jmath \omega_{\rm S} (M_{\rm B}-1) \cos \theta} \big]^T, \\
    \mathbf{a}_{\rm I}(\varphi)  &= \big[ 1, e^{\jmath \omega_{\rm S} \cos \varphi}, \dots, e^{\jmath \omega_{\rm S} (M_{\rm I}-1) \cos \varphi} \big]^T,
\end{align}
where $\omega_{\rm S} = \frac{2\pi d}{\lambda}$ with $\lambda$ denoting the wavelength.
We also define $\mathbf{h}^{{\rm UI},(q)}_{k,n,t} \in \mathbb{C}^{M_{\text{I}} \times 1}$ as the user $k$-IRS $r$ channel at the sub-carrier $n$ and symbol interval $q$ in the frame $t$, $\forall k,t$. Moreover, define $\phi_{m_\text{I},r,t}^{(q)}$ as the reflecting coefficient of the $m_\text{I}$-th RIS element with $|\phi_{m_\text{I},r,t}^{(q)}| = 1$ during the $q$-th OFDM symbol duration in the frame $t$, and $\pmb{\phi}_{r,t}^{(q)}=[\phi_{1,r,t}^{(q)},\cdots,\phi_{M_\text{I},r,t}^{(q)}]^T \in \mathbb{C}^{M_\text{I}\times 1}$. Therefore, the effective channel of the user $k$-BS channel at the $n$-th subcarrier and the $q$-th OFDM symbol interval in the frame $t$ can be defined as
\begin{small}
\begin{align}\label{equ:UE-BS_channel}
  \mathbf{h}_{k,n,t}^{(q)} 
  &=  \mathbf{h}_{k,n,t}^{{\rm UB}, (q)} + \sum_{r=1}^{R} \mathbf{G}_{r,n} \text{diag}(\pmb{\phi}_{r,t}^{(q)}) \mathbf{h}_{k,r,n,t}^{{\rm UI}, (q)}. 
\end{align}
\end{small}In the above user $k$-BS channel, the user $k$-BS link component and the user $k$-RIS $r$ link component can be written as
\begin{small}
\begin{align}
  \mathbf{h}_{k,n,t}^{{\rm UB}, (q)} &= \alpha_{k,t}^{\rm UB} \beta^{\rm UB}_{k,t} e^{-\jmath 2\pi (\Delta f (n-1) \tau^{\rm UB}_{k,t} + \Delta t (q-1) \nu_{k,t}^{\rm UB})} \mathbf{a}_{\rm B}(\theta^{\rm UB}_{k,t}), \label{equ:h_UB} \\ 
  \mathbf{h}_{k,r,n,t}^{{\rm UI}, (q)} &= \alpha_{k,r,t}^{\rm UI} \beta^{\rm UI}_{k,r,t} e^{-\jmath 2\pi (\Delta f (n-1) \tau^{\rm UI}_{k,r,t} + \Delta t (q-1) \nu_{k,r,t}^{\rm UI})} \mathbf{a}_{\rm I}(\theta^{\rm UI}_{k,r,t}), \label{equ:h_UI} 
\end{align}
\end{small}where $\alpha^{\rm UB}_{k,t} (\alpha^{\rm UI}_{k,r,t}) \in \{0,1\}$ indicates the blockage condition of the LOS user $k$-BS (user $k$-RIS $r$) path\footnote{Without loss of generality, we assume that $\alpha^{{\rm UB}}_{k,t} + \sum_{r=1}^{R}\alpha^{\rm UI}_{k,r,t} \ge 1, ~\forall k,t$ always holds, meaning that there exists at least one direct or indirect link between each user $k$ and the BS for available measurements for tracking and signal detection in each frame $t$.}; $\beta_{k,t}^{\rm UB} (\beta^{\rm UI}_{k,r,t})$ denotes the complex channel gain including the path loss and coefficient for the LOS user $k$-BS (user $k$-RIS $r$) path; $\tau_{k,t}^{\rm UB} (\tau^{\rm UI}_{k,r,t})$ and $\nu^{\rm UB}_{k,t} (\nu^{\rm UI}_{k,r,t})$ denote the delay and Doppler frequency of the LOS path between the user $k$ and the BS (RIS $r$), respectively; 
$\theta_{k,t}^{\rm UB} (\theta_{k,r,t}^{\rm UI})$ denotes the AOA of the user $k$-BS (user $k$-RIS $r$) path.
In the above channel model \eqref{equ:h_UB} and \eqref{equ:h_UI}, the delay profile, Doppler frequency, and AOA for the user $k$-BS (user $k$-RIS $r$) path during the $t$-th frame can be calculated as
\begin{align}
    &\tau_{k,t}^{\rm UB} = c_0^{-1}d_{k,t}^{\rm UB},~\forall k,t, \label{equ:delay_state} \\ 
    &\nu^{\rm UB}_{k,t} = \lambda^{-1}(\dot{\mathbf{p}}_{k,t}^{\rm U})^T \mathbf{e}^{\rm UB}_{k,t}, ~\forall k,t, \label{equ:Doppler_state} \\ 
    &\theta_{k,t}^{\rm UB} = \arccos( (\mathbf{e}^{\rm B})^T \mathbf{e}^{\rm UB}_{k,t} ), ~\forall k,t, \label{equ:DOA_state} 
\end{align}
respectively, where $d_{k,t}^{\rm UB} = ||\mathbf{p}_{k,t}^{\rm U} - \mathbf{p}^{\rm B}||_2, ~\forall k,t$ is the distance between user $k$ and the BS at the frame $t$, $\mathbf{e}^{\rm UB}_{k,t} = (\mathbf{p}^{\rm U}_{k,t} - \mathbf{p}^{\rm B})/\lambda$ is the directional vector from the BS to the user $k$, and $\mathbf{e}^{\rm B} (\mathbf{e}_{r}^{\rm I}) \in \mathbb{R}^{2 \times 1}$ is the array steering vector of the BS (RIS $r$). The parameters $\{\tau_{k,r,t}^{\rm UI},\nu_{k,r,t}^{\rm UI},\theta_{k,r,t}^{\rm UI}\}$ can be similarly derived by following \eqref{equ:delay_state} -- \eqref{equ:DOA_state}.

With the above, the received signal of the ISAC sub-block in each frame $t$ can be expressed by a fourth-order tensor as
\begin{small}
\begin{align}\label{equ:Y_tensor}
    \pmb{\mathcal{Y}}_t =&~  \sum_{k=1}^{K} w_{k,t}^{\rm UB} \Big( \mathbf{a}_{\rm B}(\theta_{k,t}^{\rm UB}) \circ \mathbf{a}_{{\rm T}_1}(\nu^{\rm UB}_{k,t}) \circ \mathbf{a}_{\rm F}(\tau^{\rm UB}_{kt}) \circ \mathbf{a}_{{\rm T}_2}(\nu^{\rm UB}_{k,t}) \Big) \notag \\
    & + \sum_{k=1}^{K} \sum_{r=1}^{R} w_{k,r,t}^{\rm UI} \Big( \mathbf{a}_{\rm B}(\xi_r) \circ \mathbf{\breve{a}}_{\rm I}(\theta^{\rm UI}_{k,r,t},\nu^{\rm UB}_{k,r,t}) \circ \mathbf{a}_{\rm F}(\tau^{\rm UIB}_{k,r,t}) \notag \\
    & \circ \mathbf{a}_{{\rm T}_2}(\nu^{\rm UI}_{k,r,t})  \Big)  + \pmb{\mathcal{Z}}_{t} \in \mathbb{C}^{M_{\rm B} \times Q_1 \times D_{\rm I} \times Q_2}, ~\forall t,
\end{align}
\end{small}where $w_{k,t}^{\rm UB} = \sqrt{P}\tilde{s}_{k,t} \alpha^{{\rm UB}}_{k,t} \beta^{\rm UB}_{k,t}$ $(w_{k,r,t}^{\rm UI} = \sqrt{P}\tilde{s}_{k,t} \alpha^{{\rm UI}}_{k,t} \beta^{\rm UI}_{k,t})$ is the effective signal along the user $k$-BS (user $k$-RIS $r$) link, $\tau^{\rm UIB}_{k,r,t} = \tau^{\rm UI}_{k,r,t} + \tau^{\rm IB}_{r}$ is the total delay of the user $k$-RIS $r$-BS link, $\pmb{\mathcal{Z}}_{t} \in \mathbb{C}^{M_{\rm B} \times Q_1 \times D_{\rm I} \times Q_2}$ is the background noise tensor with the Gaussian distribution. 
In the above equation \eqref{equ:Y_tensor}, $\breve{\mathbf{a}}_{\rm I}(\cdot)$ $\mathbf{a}_{\rm F}(\cdot)$ and $\mathbf{a}_{\rm F}(\cdot)$ are the RIS-related, frequency-related, and time-related steering vectors defined by
\begin{align}
  \breve{\mathbf{a}}_{\rm I}(\theta,\nu) &= (\pmb{\Psi}_r^T (\mathbf{a}_{\rm I}(\varphi_r)\odot\mathbf{a}_{\rm I}(\theta))) \odot \mathbf{a}_{{\rm T}_1}(\nu), \\
  \mathbf{a}_{\rm F}(\tau) &= \big[ 1, e^{-\jmath \omega_{\rm F} \tau}, \dots, e^{-\jmath \omega_{\rm F} (N_0 - 1) \tau} \big]^T, \\
  \mathbf{a}_{{\rm T}_1}(\nu)  &= \big[ 1, e^{-\jmath \omega_{{\rm T}_1} \nu},    \dots, e^{-\jmath \omega_{{\rm T}_1} (Q_1 - 1) \nu} \big]^T, \\
  \mathbf{a}_{{\rm T}_2}(\nu)  &= \big[ 1, e^{-\jmath \omega_{{\rm T}_2} \nu},    \dots, e^{-\jmath \omega_{{\rm T}_2} (Q_2 - 1) \nu} \big]^T,
\end{align}
respectively, where $\omega_{\rm F} = 2\pi \Delta N\Delta f$, $\omega_{{\rm T}_1} = 2\pi \Delta t$, and $\omega_{{\rm T}_2} = 2\pi \Delta Q\Delta t$.


\subsection{Mobility Model for Users}\label{subsec:state model}

For the mobility model, we define the state vector for all users at each frame $t$ as $\pmb{\psi}_t = [\pmb{\psi}_{1,t}^T, \dots, \pmb{\psi}_{K,t}^T]^T \in \mathbb{R}^{4K \times 1}$ with $\pmb{\psi}_{k,t} = [(\mathbf{p}_{k,t}^{\rm U})^T, (\mathbf{\dot{p}}_{k,t}^{\rm U})^T]^T \in \mathbb{R}^{4 \times 1}$. By assuming that the motion of each user follows the discrete-time state model, the state transition equation can be written as
\begin{align}\label{equ:discrete_mm}
    \pmb{\psi}_{t} = \mathbf{F} \pmb{\psi}_{t-1} + \pmb{\omega}_{t-1}, ~\forall t,
\end{align}
where the transition matrix $\mathbf{F}$ is defined as $\mathbf{F} = \mathbf{I}_{K} \otimes \mathbf{F}_0$ with
\begin{align}
    \mathbf{F}_0 &= \begin{bmatrix}
        \mathbf{I}_{2} & \Delta T \mathbf{I}_{2}  \\
        \mathbf{0} & \mathbf{I}_{2} 
    \end{bmatrix} \in \mathbb{R}^{4K \times 4K}
\end{align}
and the vector $\pmb{\omega}_{t-1} \in \mathbb{R}^{4K \times 1}$ denotes the motion noise following the Gaussian distribution $\mathcal{N}(\mathbf{0},\mathbf{Q}_{t-1})$ with $\mathbf{Q}_{t} = \text{blkdiag}(\mathbf{Q}_{1,t},\dots,\mathbf{Q}_{K,t})$.
It is assumed that the motion noise vector $\pmb{\omega}_t$ and $\pmb{\omega}_{t'}$ are mutually independent for $\forall t \ne t'$. Based on the above mobility model, we can utilize historical measurements to refine the multiuser tracking performance, which also enhances the signal detection performance.

\subsection{Problem Formulation}

The objective of this work is to jointly track the position $\mathbf{p}_{k,t}^{\rm U}$ and detect the transmit signal $\{\tilde{s}_{k,t}\}$ of each user $k$ based on received signals $\{\pmb{\mathcal{Y}}_{v}\}_{v=1}^{t}$. Such blind estimation problem is quite challenging to solve due to the complicated coupling among user states, state-related parameters, and transmit signals in the signal model \eqref{equ:Y_tensor}. To address this issue, we first propose to establish a probabilistic model for the considered system and then leverage the Bayesian estimation framework to solve the joint estimation problem. 

We first apply the geometric constraints \eqref{equ:delay_state} -- \eqref{equ:DOA_state} to build the statistical relationship between the user state and state-related parameters, which can be given by
\begin{align}
  p(\tau^{\rm UB}_{k,t}|\pmb{\psi}_{k,t}) &= \delta\left(\tau_{k,t}^{\rm UB} - c_0^{-1}d_{k,t}^{\rm UB}\right), \\
  p(\nu^{\rm UB}_{k,t}|\pmb{\psi}_{k,t}) &= \delta\left(\nu^{\rm UB}_{k,t} - (\dot{\mathbf{p}}_{k,t}^{\rm U})^T \mathbf{e}_{k,t}^{\rm UB} / \lambda \right), \\
  p(\theta_{k,t}^{\rm UB}|\pmb{\psi}_{k,t}) &= \delta\left(\theta_{k,t}^{\rm UB} - \arccos( (\mathbf{e}^{\rm B})^T \mathbf{e}^{\rm U}_{k,t})\right),
\end{align}
and the conditional probabilities for $\{\tau_{k,r,t}^{\rm UI},\nu_{k,r,t}^{\rm UI},\theta_{k,r,t}^{\rm UI}\}$ over $\{\pmb{\psi}_{k,t}\}$ can be similarly defined. By following the discrete mobility model \eqref{equ:discrete_mm}, the state evolution process can be represented by a first-order Markov process as
\begin{align}\label{equ:mm_markov}
  p(\pmb{\psi}_{k,t}|\pmb{\psi}_{k,t-1}) &= \mathcal{N}(\pmb{\psi}_{k,t};\mathbf{F}_0\pmb{\psi}_{k,t-1},\mathbf{Q}_{k,t-1}), ~\forall k,t.
\end{align}

Denote $\mathbf{s}_{t} = [\tilde{s}_{1,t}, \dots, \tilde{s}_{K,t}]$ as the transmit signal vector and $\pmb{\vartheta}^{\rm UB}_{t} = [(\pmb{\vartheta}^{\rm UB}_{1,t})^T, \dots, (\pmb{\vartheta}^{\rm UB}_{K,t})^T]^T$ $(\pmb{\vartheta}^{\rm UI}_{t} = [(\pmb{\vartheta}^{\rm UI}_{1,t})^T, \dots, (\pmb{\vartheta}^{\rm UI}_{K,t})^T]^T)$ as the state-related parameter vector for user-BS (user-RIS) links, where $\pmb{\vartheta}^{\rm UB}_{k,t} = [\alpha^{\rm UB}_{k,t}, \tau^{\rm UB}_{k,t}, \nu^{\rm UB}_{k,t}, \theta^{\rm UB}_{k,t}]^T$,  $\pmb{\vartheta}^{\rm UI}_{k,t}  = [(\pmb{\vartheta}^{\rm UI}_{k,1,t})^T, \dots, (\pmb{\vartheta}^{\rm UI}_{k,R,t})^T]^T$, and $\pmb{\vartheta}^{\rm UI}_{k,r,t} = [\alpha^{\rm UI}_{k,r,t}, \tau^{\rm UI}_{k,r,t}, \nu^{\rm UI}_{k,r,t}, \theta^{\rm UI}_{k,r,t}]^T$.
By following \eqref{equ:Y_tensor}, the conditional probability of the received signal in the ISAC sub-block of each frame $t$ over the state-related parameter of users can be denoted by $p(\pmb{\mathcal{Y}}_{t}|\pmb{\vartheta}^{\rm UB}_{t}, \pmb{\vartheta}^{\rm UI}_{t}, \mathbf{s}_{n,t}^{(q)})$ in the forms of the complex Gaussian distribution.
Therein, the probabilistic density function (PDF) of the transmit signal of each user $k$ is denoted by $p(\tilde{s}_{k,t}) = \mathcal{P}_0(\tilde{s}_{k,t}), ~\forall k,t$.
Therefore, the joint probability can be represented by
\begin{small}
\begin{align}\label{equ:joint_prob}
    &p(\{\pmb{\mathcal{Y}}_{t},\pmb{\vartheta}^{\rm UB}_{t}, \pmb{\vartheta}^{\rm UI}_{t}, \pmb{\psi}_{t}, \mathbf{s}^{(q)}_{t}\}) = \prod_{t} p(\pmb{\mathcal{Y}}_{t}|\pmb{\vartheta}^{\rm UB}_{t}, \pmb{\vartheta}^{\rm UI}_{t}, \mathbf{s}_{t}) \notag \\
    &\times \bigg( \prod_{k=1}^{K} 
    p(\pmb{\vartheta}^{\rm UB}_{k,t}, \pmb{\vartheta}^{\rm UI}_{k,t}| \pmb{\psi}_{k,t})  p(\pmb{\psi}_{k,t}|\pmb{\psi}_{k,t-1}) \mathcal{P}_0(\tilde{s}_{k,t}) \bigg),
\end{align}
\end{small}where $p(\pmb{\psi}_{k,1}|\pmb{\psi}_{k,0}) = p(\pmb{\psi}_{k,1})$ denotes the prior information of the user state at the first frame derived from the initial estimation. Based on the above probabilistic model, we can follow the Bayes' rule to derive the posterior probability of the position and transmit signal of each user $k$ in the frame $t$ over the received signals in frames $v \in \{1,\dots,t\}$ as 
\begin{small}
\begin{align}
    p(\pmb{\psi}_{k,t}|\{\pmb{\mathcal{Y}}_v\}_{v = 1}^{t}) &= \int_{\setminus \pmb{\psi}_{k,t}} \frac{p(\{\pmb{\mathcal{Y}}_v,\pmb{\vartheta}^{\rm UB}_{v}, \pmb{\vartheta}^{\rm UI}_{v}, \pmb{\psi}_{v}, \mathbf{s}_{v}\}_{v = 1}^{t})}{p(\{\pmb{\mathcal{Y}}_{v}\}_{v = 1}^{t})}, \label{equ:postprob_state}  \\
    p(\tilde{s}_{k,t}|\{\pmb{\mathcal{Y}}_v\}_{v = 1}^{t}) &= \int_{\setminus \tilde{s}_{k,t}}\frac{ p(\{\pmb{\mathcal{Y}}_{v},\pmb{\vartheta}^{\rm UB}_{v}, \pmb{\vartheta}^{\rm UI}_{v}, \pmb{\psi}_{v}, \mathbf{s}_{v}\}_{v = 1}^{t})}{p(\{\pmb{\mathcal{Y}}_{v}\}_{v = 1}^{t})}. \label{equ:postprob_signal}
\end{align}
\end{small}After that, we can apply the MAP criterion to perform the online estimation of the user states and transmit signal of each user from all the received signals $\{\mathcal{Y}_v\}_{v=1}^{t}$. However, it is intractable to derive explicit expressions on these posterior probabilities \eqref{equ:postprob_state} -- \eqref{equ:postprob_signal}, due to high-dimensional integrals. Although the popular KF-based methods \cite{Einicke_1999_TSP} can also be utilized for the tracking problem, their performance suffers severe deterioration in the scenario with limited tracking resources. As such, this work resorts to the Bayesian approximate inference framework, including message passing and variational message passing techniques to design a computationally efficient joint estimation algorithm.

\section{Joint Multiuser Tracking and Signal Detection Algorithm}

In this section, we propose an HVMP algorithm for the joint multiuser tracking and signal detection problem. 

\begin{figure}[t]
  \centering
  \includegraphics[width=.42\textwidth]{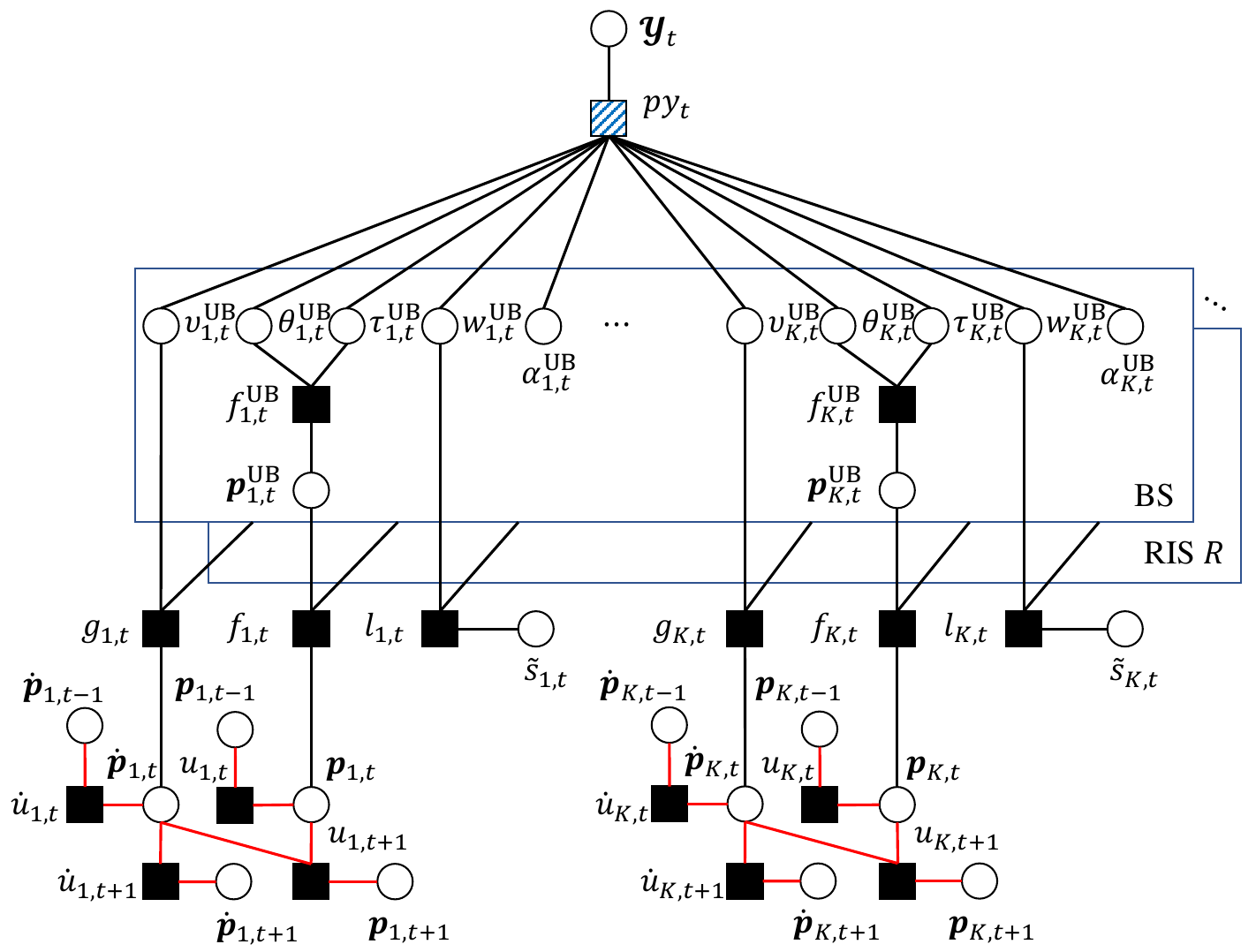}
  \vspace{-0.3cm}
  \caption{Factor graph of the probabilistic model.}\label{Fig:factor_graph}
  \vspace{-0.6cm}
\end{figure}

\subsection{Graphical Representation}
Before deriving message passing equations in the proposed HVMP algorithm, we first establish the graphical representation of the factor graph based on the joint probability \eqref{equ:joint_prob} as shown in Fig. \ref{Fig:factor_graph}. Therein, squares and white circles denote factor nodes and variable nodes, respectively. Moreover, we propose to perform bidirectional message passing over (black) edges in the factor graph, and perform only forward message passing over (red) edges for state update, due to the online estimation principle. In particular, the VMP is performed along edges connected with the blue-veined square, while the standard MP is realized along the other edges. The details of the proposed algorithm are given in the next subsection. 

\subsection{Hybrid Variational Message Passing}

In general, the HVMP algorithm integrates VMP and MP techniques to update all the extrinsic messages with tractable expressions, thus realizing computationally efficient online estimation for user states and transmit signals.
In the following, we first introduce the bidirectional message passing equations sequentially and then give the forward message passing equations for the online estimation. In particular, we consider modeling distributions with states by Gaussian distributions, while von Misés (VM) distributions \cite{Badiu_2017_TSP} are used for the state-related parameters in the proposed algorithm. Due to page limit, detailed derivations of the following messages are omitted.

\subsubsection{Messages between $\dot{\mathbf{p}}_{k,t}$ and $g_{k,t}$}
The bidirectional messages in the $j$-th iteration can be given by
\begin{small}
\begin{align}
    \varepsilon^j_{\dot{\mathbf{p}}_{k,t} \rightarrow g_{k,t}} &= \mathcal{N}(\dot{\mathbf{p}}_{k,t}; \dot{\overrightarrow{\mathbf{p}}}_{k,t}, \dot{\overrightarrow{\mathbf{Q}}}_{k,t}), ~\forall j, \\
    \varepsilon^j_{\dot{\mathbf{p}}_{k,t} \leftarrow g_{k,t}}   & \varpropto \int p(\nu^{\rm UB}_{k,t},\{\nu^{\rm UI}_{k,r,t}\}|\dot{\mathbf{p}}_{k,t}) \varepsilon^j_{g_{k,t} \leftarrow \nu^{\rm UB}_{k,t}} \prod_{r=1}^{R} \varepsilon^j_{g_{k,t} \leftarrow \nu^{\rm UI}_{k,r,t}} \notag \\
     &= \mathcal{N}(\dot{\mathbf{p}}_{k,t}; \dot{\overleftarrow{\mathbf{p}}}^{j}_{k,t}, \dot{\overleftarrow{\mathbf{Q}}}^{j}_{k,t}), ~\forall j,
\end{align}
\end{small}where $\dot{\overrightarrow{\mathbf{p}}}_{k,t}$ and $\dot{\overrightarrow{\mathbf{Q}}}_{k,t}$ are the mean and covariance of the velocity from the online estimation in the last frame $t-1$, $\dot{\overleftarrow{\mathbf{p}}}^{j}_{k,t}$ and $\dot{\overleftarrow{\mathbf{Q}}}^{j}_{k,t}$ are the updated ones from the Doppler frequency estimation in the $j$-th iteration. The exact expression of $\dot{\overleftarrow{\mathbf{p}}}^{j}_{k,t}$ and $\dot{\overleftarrow{\mathbf{Q}}}^{j}_{k,t}$ is hard to be derived due to the complicate relationship between Doppler frequencies and the user velocity. Thus, we resort to the Gaussian-Newton method and Taylor series expansion to obtain $\dot{\overleftarrow{\mathbf{p}}}^{j}_{k,t}$ and $\dot{\overleftarrow{\mathbf{Q}}}^{j}_{k,t}$ via Gaussian approximations. 

\subsubsection{Messages between $\mathbf{p}_{k,t}$ and $f_{k,t}$}
The bidirectional messages in the $j$-th iteration are derived by
\begin{small}
\begin{align}
    \varepsilon^j_{\mathbf{p}_{k,t} \rightarrow f_{k,t}} &= \mathcal{N}(\mathbf{p}_{k,t}; \overrightarrow{\mathbf{p}}_{k,t}, \overrightarrow{\mathbf{Q}}_{k,t}), ~\forall j, \\
    \varepsilon^j_{\mathbf{p}_{k,t} \leftarrow f_{k,t}}   & \varpropto \int p(\mathbf{p}^{\rm UB}_{k,t},\{\mathbf{p}^{\rm UI}_{k,r,t}\}|\mathbf{p}_{k,t}) \varepsilon^j_{f_{k,t} \leftarrow \mathbf{p}^{\rm UB}_{k,t}} \prod_{r=1}^{R} \varepsilon^j_{f_{k,t} \leftarrow \mathbf{p}^{\rm UI}_{k,r,t}} \notag \\
     &= \mathcal{N}(\mathbf{p}_{k,t}; \overleftarrow{\mathbf{p}}^{j}_{k,t}, \overleftarrow{\mathbf{Q}}^{j}_{k,t}), ~\forall j, \label{equ:mp_fusion_p}
\end{align}
\end{small}
where 
\begin{small}
\begin{align}
  \overleftarrow{\mathbf{p}}^j_{k,t} =&~ (\overleftarrow{\mathbf{Q}}^{j}_{k,t})^{-1} [ (\overleftarrow{\mathbf{Q}}^{{\rm UB}, j}_{k,t})^{-1} \overleftarrow{\mathbf{p}}^{{\rm UB}, j}_{k,t} + \sum_{r=1}^{R} (\overleftarrow{\mathbf{Q}}^{{\rm UI}, j}_{k,r,t})^{-1} \overleftarrow{\mathbf{p}}^{{\rm UI}, j}_{k,r,t} ] , \notag \\
  \overleftarrow{\mathbf{Q}}^{j}_{k,t} =&~ \Big( (\overleftarrow{\mathbf{Q}}^{{\rm UB}, j}_{k,t} )^{-1} + \sum_{r=1}^{R}  (\overleftarrow{\mathbf{Q}}^{{\rm UI}, j}_{k,r,t} )^{-1} \Big)^{-1} , \notag
\end{align}
\end{small}and $\overrightarrow{\mathbf{p}}_{k,t}$ ($\overrightarrow{\mathbf{Q}}_{k,t}$) is the mean (covariance) of the position from the online estimation in the last frame $t-1$. 

\subsubsection{Messages between $\tilde{s}_{k,t}$ and $l_{k,t}$}
The message from $\tilde{s}_{k,t}$ to $l_{k,t}$ is obtained based on the prior distribution of the transmit signal, i.e., $\varepsilon^j_{\tilde{s}_{k,t} \rightarrow l_{k,t}} = \mathcal{P}_0(\tilde{s}_{k,t})$, and the message from $l_{k,t}$ to ${\tilde{s}}_{k,t}$ is derived by
\begin{small}
\begin{align}
  \varepsilon^{j}_{\tilde{s}_{k,t} \leftarrow l_{k,t}} \varpropto&~ \int p(w^{\rm UB}_{k,t}, \{w^{\rm UI}_{k,r,t}\}| \tilde{s}_{k,t})  \varepsilon^j_{l_{k,t} \leftarrow w^{\rm UB}_{k,t}} \prod_{r=1}^{R} \varepsilon^j_{l_{k,t} \leftarrow w^{\rm UI}_{k,r,t}} \notag \\
  =&~ \mathcal{CN}(\tilde{s}_{k,t}; \overleftarrow{s}^{j}_{k,t},\overleftarrow{\rho}^{j}_{k,t}),  ~\forall j,
\end{align}
\end{small}where
\begin{small}
\begin{align}
  \overleftarrow{s}^{j}_{k,t} &= (\overleftarrow{\rho}^{j}_{k,t})^{-1} ( (\overleftarrow{\varpi}^{{\rm UB},j}_{k,t})^{-1}\overleftarrow{w}^{{\rm UB},j}_{k,t} + \sum_{r=1}^{R} (\overleftarrow{\varpi}^{{\rm UI},j}_{k,r,t})^{-1}\overleftarrow{w}^{{\rm UI},j}_{k,r,t}  ), \notag \\
  \overleftarrow{\rho}^{j}_{k,t} &= \Big((\overleftarrow{\varpi}^{{\rm UB},j}_{k,t})^{-1} +  \sum_{r=1}^{R} (\overleftarrow{\varpi}^{{\rm UI},j}_{k,r,t})^{-1} \Big)^{-1}. \notag
\end{align}
\end{small}

\subsubsection{Messages between $g_{k,t}$ and $\nu^{\rm UB}_{k,t}$ ($\nu^{\rm UI}_{k,r,t}$)}
The bidirectional message passing equations can be written as
\begin{small}
\begin{align}
  \varepsilon^{j}_{g^{\rm UB}_{k,t} \rightarrow \nu^{\rm UB}_{k,t}} &\varpropto \int p(\nu^{\rm UB}_{k,t},\{\nu^{\rm UI}_{k,r,t}\}|\dot{\mathbf{p}}_{k,t}) \prod_{r=1}^{R} \varepsilon^j_{g_{k,t} \leftarrow \nu^{\rm UI}_{k,r,t}} \notag \\
  &= \mathcal{VM}(-\varpi_{\rm T} \nu^{\rm UB}_{k,t}; \overrightarrow{\mu}^{\nu, j}_{k,t}, \overrightarrow{\kappa}^{\nu, j}_{k,t}), \\
  \varepsilon^{j}_{g^{\rm UB}_{k,t} \leftarrow \nu^{\rm UB}_{k,t}}   &= \mathcal{VM}(-\varpi_{\rm T} \nu^{\rm UB}_{k,t}; \overleftarrow{\mu}^{\nu, j}_{k,t}, \overleftarrow{\kappa}^{\nu, j}_{k,t}),
\end{align}
\end{small}where $\mathcal{VM}(\nu;\cdot,\diamond)$ denotes the VM distribution, the $\overrightarrow{\mu}^{\nu, j}_{k,t}$ ($\overrightarrow{\kappa}^{\nu, j}_{k,t}$) are also derived by the Gauss-Newton method, and $\varepsilon^{j}_{g^{\rm UB}_{k,t} \leftarrow \nu^{\rm UB}_{k,t}}$ is derived from the VMP part. 

\subsubsection{Messages between $f_{k,t}$ and $\mathbf{p}^{\rm UB}_{k,t}$ ($\mathbf{p}^{\rm UI}_{k,r,t}$)}
Due to the Gaussian message passing along the edges connected with node $f_{k,t}$, we can simply derive the message $\varepsilon^{j}_{f_{k,t} \rightarrow \mathbf{p}^{\rm UB}_{k,t}}$ ($\varepsilon^{j}_{f_{k,t} \rightarrow \mathbf{p}^{\rm UI}_{k,r,t}}$) similar to \eqref{equ:mp_fusion_p}, and the message $\varepsilon^{j}_{f_{k,t} \leftarrow \mathbf{p}^{\rm UB}_{k,t}}$ ($\varepsilon^{j}_{f_{k,t} \leftarrow \mathbf{p}^{\rm UI}_{k,r,t}}$) can be obtained as $\mathcal{N}(\mathbf{p}^{\rm UB}_{k,t}; \overleftarrow{\mathbf{p}}^{{\rm UB},j}_{k,t},\overleftarrow{\mathbf{Q}}^{{\rm UB},j}_{k,t})$,
where the mean and covariance matrix are derived based on second-order Taylor series expansion and Gaussian approximation based on the updated messages on $\theta^{\rm UB}_{k,t}$ and $\tau^{\rm UB}_{k,t}$ in the part of VMP.

\subsubsection{Messages between $l_{k,t}$ and $w^{\rm UB}_{k,t}$ ($w^{\rm UI}_{k,r,t}$)}
The bidirectional messages can be calculated by
\begin{align}
  \varepsilon^{j}_{l_{k,t} \rightarrow w^{\rm UB}_{k,t}} &\varpropto \int p(w^{\rm UB}_{k,t}, \{w^{\rm UI}_{k,r,t}\}| \tilde{s}_{k,t}) \prod_{r=1}^{R} \varepsilon^j_{l_{k,t} \leftarrow w^{\rm UI}_{k,r,t}}, \label{equ:l_r_w} \\
  \varepsilon^{j}_{l_{k,t} \leftarrow w^{\rm UB}_{k,t}}   &= \mathcal{CN}(w^{\rm UB}_{k,t}; \overleftarrow{w}^{{\rm UB},j}_{k,t} ,  \overleftarrow{\varpi}^{{\rm UB},j}_{k,t}),  \label{equ:l_l_w}
\end{align}
where $\overleftarrow{w}^{{\rm UB},j}_{k,t}$ and $\overleftarrow{\varpi}^{{\rm UB},j}_{k,t}$ are derived from the variational message passing. Messages over the variable $w^{\rm UI}_{k,r,t}$ can be similarly derived by following \eqref{equ:l_l_w} and \eqref{equ:l_r_w}.

\subsubsection{Messages between $f^{\rm UB}_{k,t}$ and $\theta^{\rm UB}_{k,t}$ (between $f^{\rm UI}_{k,t}$ and $\theta^{\rm UI}_{k,r,t}$) and messages between $f^{\rm UB}_{k,t}$ and 
$\tau^{\rm UB}_{k,t}$ (between $f^{\rm UI}_{k,t}$ and $\tau^{\rm UI}_{k,r,t}$)}
By following the derivation of the message $\varepsilon^{j}_{g^{\rm UB}_{k,t} \rightarrow \nu^{\rm UB}_{k,t}}$, we also leverage the second-order Taylor series expansion and VM approximation to calculate messages $\varepsilon^{j}_{f^{\rm UB}_{k,t} \rightarrow \theta^{\rm UB}_{k,t}}$ and $\varepsilon^{j}_{f^{\rm UB}_{k,t} \rightarrow \tau^{\rm UB}_{k,t}}$, where is represented in the form of VM distributions. Alternatively, messages $\varepsilon^{j}_{f^{\rm UB}_{k,t} \leftarrow \theta^{\rm UB}_{k,t}}$ and $\varepsilon^{j}_{f^{\rm UB}_{k,t} \leftarrow \tau^{\rm UB}_{k,t}}$ are obtained from the variational message passing. The messages for user-RIS links can be similarly obtained by following the above updating principle.

\subsubsection{Messages between $py_{t}$ and $\pmb{\vartheta}^{\rm UB}_{k,t}$ $(\pmb{\vartheta}^{\rm UI}_{k,r,t})$ via variational message passing}
In this part, the variational message passing is utilized to update the posterior probabilities of all the state-related parameters. The key idea of variational message passing is to approximate the joint distributions of these variables by the surrogate probability, which has the minimal Kullback-Leibler (KL) divergence to the exact joint probability. To make the probability optimization tractable, the surrogate probability can usually be factorized into the product of a set of independent probabilities, each of which only depends on the individual variable. As such, the coordinate descent method is usually applied to optimize each variable for minimizing the KL divergence. Specifically, we adopt the VM distribution to approximate the posterior distribution of each position-related variable. After manipulations, the objective function of the optimization subproblem over each variable $\vartheta \in \{\theta^{\rm UB}_{k,t}, \tau^{\rm UB}_{k,t}, \nu^{\rm UB}_{k,t}\}_{k=1}^{K} \cup \{\theta^{\rm UI}_{k,r,t}, \tau^{\rm UI}_{k,r,t}, \nu^{\rm UI}_{k,r,t}\}_{k=1,r=1}^{K,R}$ can be formulated as
\begin{align}\label{equ:opt_BCD_variable}
    \ln \varepsilon^{j,\iota}(\vartheta) = \ln \varepsilon^{j}_{u \rightarrow \vartheta} + \Re\{ (\pmb{\eta}^{j,\iota}_{\vartheta})^H \mathbf{a}_{\rm i}(\vartheta) \},
\end{align}
where ${\rm i} \in \{\rm S, I, F, T\}$, $u$ is the factor node connected with variable node $\vartheta$, and $\iota$ denotes the inner iteration index of the variational message passing. This unconstrained optimization problem \eqref{equ:opt_BCD_variable} is solved by the Heuristic Algorithm 2 in \cite{Badiu_2017_TSP} via coordinate descent. Then, the parameters $\{w^{\rm UB}_{k,t},w^{\rm UB}_{k,1,t},\dots,w^{\rm UB}_{k,R,t}\}$ can be derived based on the minimum mean square error (MMSE) estimation.

In this part, we also propose to update the posterior distribution $\delta(\alpha^{\rm UB}_{k,t} - \hat{\alpha}^{\rm UB}_{k,t})$ of the indicator variable $\alpha^{\rm UB}_{k,t}$ ($\alpha^{\rm UI}_{k,r,t}$) by utilizing the greedy iterative search method in \cite{Badiu_2017_TSP}.
In each outer iteration of the HVMP algorithm, we perform VMP in $I_{\rm inn}$ iterations to converge to a stationary point.

\subsubsection{Forward messages}
After the state estimation in the current frame $t$ converges, we can update the forward messages to the frame $t+1$ for online user tracking and signal detection. These forward messages can be expressed by
\begin{align}
    \varepsilon_{\dot{u}_{k,t+1} \rightarrow \dot{\mathbf{p}}_{k,t+1}} &= \varepsilon_{\dot{\mathbf{p}}_{k,t} \rightarrow \dot{u}_{k,t+1}} = \varepsilon_{\dot{\mathbf{p}} \rightarrow u_{k,t+1}} \notag \\
    &\varpropto \varepsilon_{\dot{u}_{k,t} \rightarrow \dot{\mathbf{p}}_{k,t}} \varepsilon^{\infty}_{\dot{\mathbf{p}}_{k,t} \leftarrow g_{k,t}}, \label{equ:mp_forward_vel} \\
    \varepsilon_{\mathbf{p}_{k,t} \rightarrow u_{k,t+1}} &\varpropto \varepsilon_{u_{k,t} \rightarrow \mathbf{p}_{k,t}} \varepsilon_{\mathbf{p}_{k,t} \leftarrow f_{k,t}}, \label{equ:mp_fusion_pos}   \\
    \varepsilon_{u_{k,t+1} \rightarrow \mathbf{p}_{k,t+1}} &\varpropto \varepsilon_{\dot{\mathbf{p}} \rightarrow u_{k,t+1}} \varepsilon_{\mathbf{p}_{k,t} \rightarrow u_{k,t+1}}. \label{equ:mp_forward_pos}
\end{align}
Since Gaussian-type distributions are contained in all the above messages \eqref{equ:mp_forward_vel} -- \eqref{equ:mp_forward_pos}, explicit expressions for distributions of user states are easy to derive and thus omitted here.

In summary, the proposed algorithm iteratively performs bidirectional message passing to update user states and state-related parameters in each frame $t$ based on the prior information from the previous frame $t-1$, and then utilizes the converged state estimation result to derive the prior information of user states for the next frame $t+1$, which realizes the online tracking and signal detection for all moving users.

\section{Simulation Results}

\begin{figure}[t]
  \centering
  \includegraphics[width=.48\textwidth]{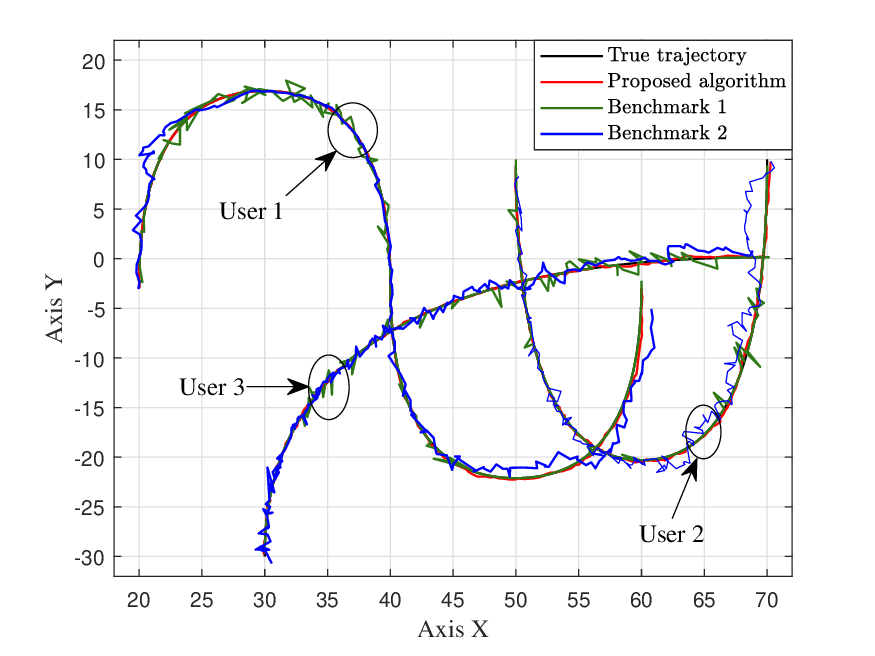}
  \vspace{-0.2cm}
  \caption{The trajectory of tracking results under receive SNR $= -10$ dB.}\label{Fig:traj}
  \vspace{-0.4cm}
\end{figure}

This section provides simulation results to verify the effectiveness of the proposed algorithm for the RIS-assisted ISAC system. In the system setup, we consider that there are one BS deployed at $\mathbf{p}^{\rm B} = [0,0]$, $R=2$ RISs deployed at $\mathbf{p}^{\rm I}_{1} = [20,40]$ and $\mathbf{p}^{\rm I}_{2} = [45,30]$, and $K = 3$ users with LOS blockage to the BS in large probability of $0.95$, which move in the square region with the center $[45,-5]$ and the side length $50$ meter. We also consider that average velocities of these three users are $||\dot{\mathbf{p}}^{\rm U}_1||_2 = 40$ m/s, $||\dot{\mathbf{p}}^{\rm U}_2||_2 = 30$ m/s, $||\dot{\mathbf{p}}^{\rm U}_3||_2 = 15$ m/s. The Gaussian signal is adopted for the uplink transmission of each user and the random phase profile \cite{Zhu_2025_arxiv} is adopted for each RIS. We consider $M_{\rm B} = 6$, $M_{\rm I} = 64$, and $N\Delta f = 10$ MHz with $N = 12$ and $\mathcal{N} = \{1,2,\dots,12\}$. The tracking interval is set as $\Delta T = 0.02$ seconds and the ISAC sub-block is set with $Q_1 = 12$, $Q_2 = 12$, and $\Delta Q = 200$. For comparison, we adopt two representative methods as the benchmark: 
\begin{itemize}
    \item \textbf{Benchmark 1:} MUSIC is utilized for the state-related parameter estimation and then KF is applied to refine the state estimation; 
    \item \textbf{Benchmark 2:} The Bayesian inference method in \cite{Teng_2023_JSAC} tracks the position of each user while neglecting the velocity estimation.
\end{itemize}
The receive signal-to-noise ratio is defined by SNR = $10\log_{10}(||\pmb{\mathcal{Y}}_{t}||_F^2/||\pmb{\mathcal{Z}}_{t}||_F^2)$. All the simulations are averaged over $100$ channel realizations.

We first give an example of the tracking trajectory derived by the proposed algorithm and benchmarks in Fig. \ref{Fig:traj}. It is observed that tracked trajectories of the proposed algorithm are always approaching true trajectories, while these two benchmarks usually have more fluctuating trajectories. Moreover, tracked trajectories of Benchmark 2 usually have larger deviations from the true trajectories, since it ignores the Doppler effect in the high-speed scenario. Therefore, the proposed algorithm can outperform both of these two representative methods.

\begin{figure}[t]
  \centering
  \includegraphics[width=.48\textwidth]{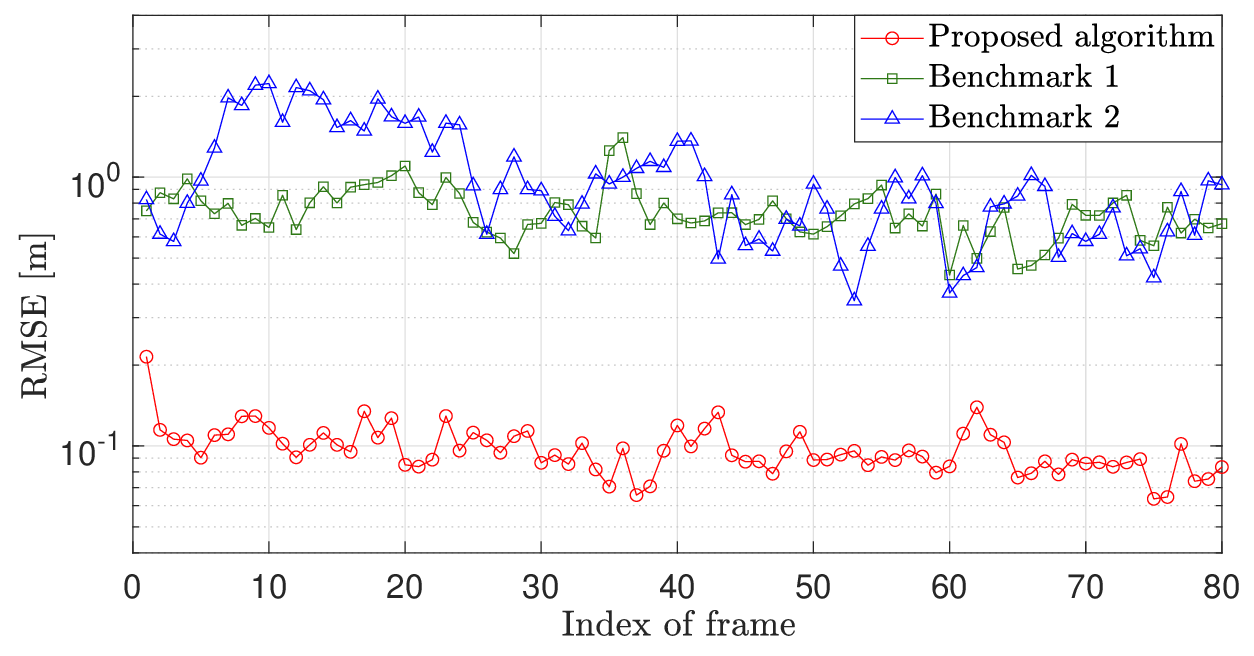}
  \vspace{-0.4cm}
  \caption{The tracking RMSE in each frame under receive SNR $= -10$ dB.}\label{Fig:RMSE_frame}
  \vspace{-0.4cm}
\end{figure}

\begin{figure}[t]
  \centering
  \includegraphics[width=.48\textwidth]{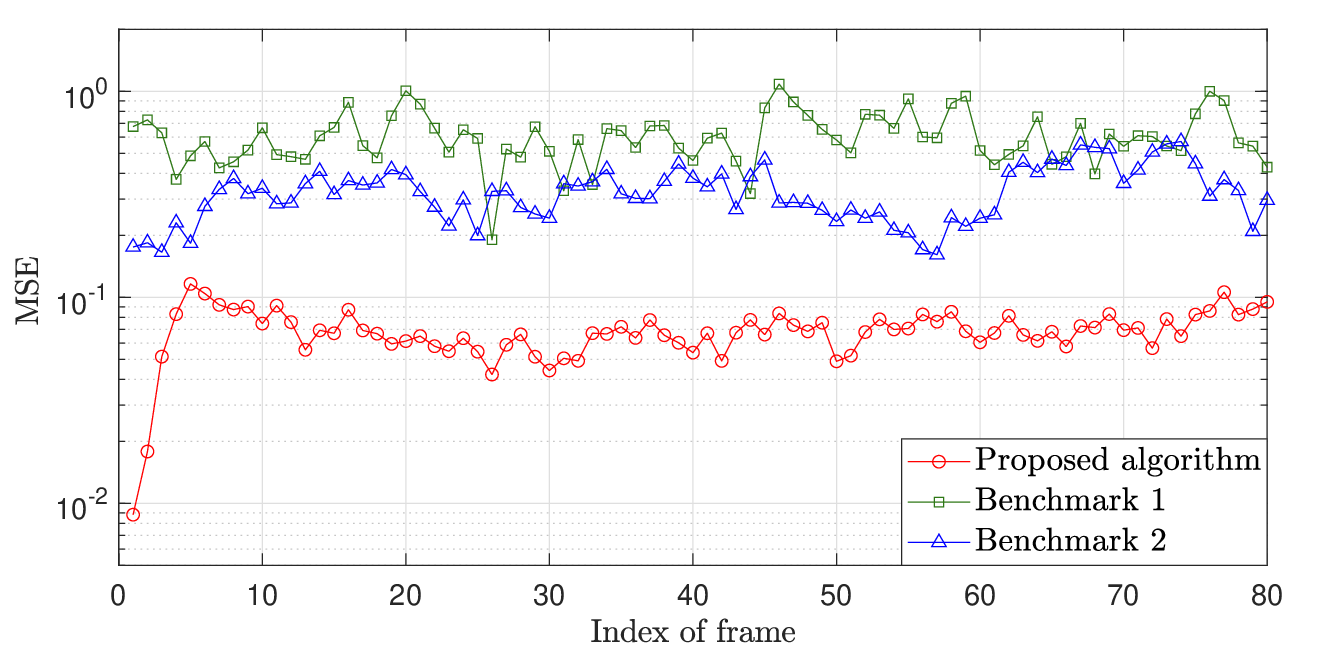}
  \vspace{-0.4cm}
  \caption{The signal detection performance in each frame under receive SNR $= -10$ dB.}\label{Fig:MSE_signal}
  \vspace{-0.4cm}
\end{figure}

Next, the tracking and signal detection performance in terms of the root mean square error (RMSE) of the estimated positions and the MSE of the estimated signals in the tracking procedure is evaluated in Fig. \ref{Fig:RMSE_frame} and \ref{Fig:MSE_signal}, respectively. We can find that the proposed algorithm can significantly outperform the two benchmarks for the tracking and signal detection problem since the proposed algorithm can better exploit the statistical interdependence in the signal model. It is also revealed that the proposed algorithm is suitable for tracking and signal detection of multiple users with high mobility in ISAC systems.

\section{Conclusions}

This work studied the joint multiuser tracking and signal detection problem in the RIS-assisted ISAC systems. Under the established probabilistic signal model, we proposed an effective and efficient HVMP algorithm to estimate user states (position and velocity) and transmit signals in an online manner. Simulation results validated that the proposed algorithm can realize high-accuracy tracking and signal detection performance for users with high-speed mobility.

\bibliographystyle{IEEEtran}
\bibliography{ref}

\end{document}